\documentclass[pra,epsfig,graphics,twocolumn,floatfix,showpacs]{revtex4}
\usepackage{amsmath,amsfonts,amssymb,graphicx,color,times,bbm,psfrag,dsfont,hyperref}
\usepackage[english]{babel}
\usepackage[latin1]{inputenc}
\usepackage[T1]{fontenc}

\newif\ifhyper
\hypertrue
\ifhyper
\hypersetup{
   citecolor = {green},
   colorlinks = {true}, 
   urlcolor = {blue} 
}
\fi

\newcommand{\beq}{\begin{equation}}
\newcommand{\eeq}{\end{equation}}
\newcommand{\beqa}{\begin{eqnarray}}
\newcommand{\eeqa}{\end{eqnarray}}
\newcommand{\ket} [1] {\vert #1 \rangle}
\newcommand{\bra} [1] {\langle #1 \vert}
\newcommand{\braket}[2]{\langle #1 | #2 \rangle}

\newcommand{\tr}{\mathop{\mathrm{tr}}}

\usepackage{graphicx}
\usepackage[mathscr]{eucal}
\usepackage{amsmath}
\usepackage{amssymb}
\usepackage{epstopdf}
\usepackage{epsfig}
\def\one{\ensuremath{\hbox{$\mathrm I$\kern-.6em$\mathrm 1$}}}
\def\tr{ \mbox{tr}}

\begin{document}

\title{The iTEBD algorithm beyond unitary evolution}

\author{R. Or\'us}
\affiliation{School of Physical Sciences, The University of Queensland, QLD 4072, Australia}

\author{G. Vidal}
\affiliation{School of Physical Sciences, The University of Queensland, QLD 4072, Australia}

\begin{abstract}
The infinite \emph{time-evolving block decimation} (iTEBD) algorithm [Phys. Rev. Lett. 98, 070201 (2007)] allows to simulate unitary evolution and to compute the ground state of one-dimensional quantum lattice systems in the thermodynamic limit. Here we extend the algorithm to tackle a much broader class of problems, namely the simulation of arbitrary one-dimensional evolution operators that can be expressed as a (translationally invariant) tensor network.
Relatedly, we also address the problem of finding the dominant eigenvalue and eigenvector of a one-dimensional transfer matrix that can be expressed in the same way. New applications include the simulation, in the thermodynamic limit, of open (i.e. master equation) dynamics and thermal states in 1D quantum systems, as well as calculations with partition functions in 2D classical systems, on which we elaborate. The present extension of the algorithm also plays a prominent role in the infinite \emph{projected entangled-pair states} (iPEPS) approach to infinite 2D quantum lattice systems.
\end{abstract}


\maketitle

\section{Introduction}

The development of numerical methods to explore the properties of strongly correlated many-body systems remains one of the most challenging problems in computational physics. In recent years, 
increasing attention has been paid to
algorithms that express the state of the system as a tensor network.
For instance, for quantum systems on a 1D lattice, a \emph{matrix product state} (MPS) \cite{MPS} represents the system's wave function in  the \emph{density matrix renormalization group} (DMRG) 
algorithm to compute ground states \cite{DMRG}, and in the \emph{time-evolving block decimation} (TEBD) algorithm to simulate time evolution \cite{TEBD}. Similarly, the \emph{tensor product state} (TPS) \cite{TPS} and the \emph{projected entangled-pair state} (PEPS) \cite{PEPS} have been proposed to accomplish those tasks in 2D lattices, whereas the \emph{multi-scale entanglement renormalization ansatz} (MERA) \cite{MERA} is specially suited to describe systems at criticality or with topological order. Finally, tensor networks can also be used to encode and manipulate the partition function of 2D classical lattice systems \cite{TMRG,CTMRG,TRG}.

The computational cost of a simulation using tensor network algorithms is roughly proportional to the size of the lattice. However, when the system is invariant under translations, this cost can be made independent of the system's size. The
infinite TEBD (iTEBD) algorithm [7] exploits this fact to simulate
unitary evolution and compute the ground state of a 1D
quantum system in the limit of an infinite lattice.
The key idea is to encode the wave function in an infinite MPS (iMPS) made of a small number of tensors that are repeated indefinitely and, importantly, to maintain the iMPS in its canonical form during the whole simulation. As a result, bulk properties of 1D quantum systems are computed directly in the thermodynamic limit, circumventing more costly and less accurate approaches based on finite-size scaling. Other algorithms, such as the \emph{power wave function renormalization group} (PWFRG) \cite{PWFRG} or infinite DMRG (iDMRG) \cite{iDMRG} also compute the ground state of infinite systems.

A major limitation of the iTEBD algorithm is that it can only address unitary evolution [as explained in Sect. III, the computation of ground states with imaginary time evolution is a lucky exception]. Thus, the simulation of more general types of evolution, such as master equation evolution in a dissipative system or imaginary time evolution to compute thermal states, is still restricted to finite systems \cite{ZV,VGC}. The reason lies in the fact that only unitary evolution preserves the canonical form of the iMPS. The latter is essential in order to keep truncation errors small during the simulation. Indeed, in the absence of the canonical form, truncation errors accumulate unnecessarily fast and ruin the simulation in places where an efficient iMPS description would otherwise still be feasible. 
 
In this paper we explain how to overcome such shortcoming. First we describe how to compute the canonical form of an iMPS. Then we present an extension of the iTEBD algorithm that is able to simulate a much wider class of evolution. Namely, it simulates the action on an iMPS of any transformation that can be expressed as a translationally invariant tensor network. This includes, as particular cases, evolution in imaginary time or according to a master equation. We also explain how to use the algorithm to compute the dominant eigenvalue and eigenvector \cite{dominant} of any one-dimensional transfer matrix that decomposes as a translationally invariant tensor network.  As an application of this, we explain how to extract correlators and local observables from the partition function of a 2D classical system. Finally, the extended version presented in this work plays a prominent role in the \emph{infinite} PEPS (iPEPS) algorithm to simulate evolution and compute ground states in infinite 2D quantum lattice systems \cite{iPEPS}, as well as in certain implementations of MERA algorithms \cite{MERA}.

We emphasize that other algorithms can be used to address infinite systems, and that they are also based on or related to computing the dominant eigenvalue and eigenvector of a one-dimensional transfer matrix. This is the case, for instance, of the PWFRG and iDMRG algorithms \cite{PWFRG,iDMRG} to compute ground states in 1D quantum systems and the transfer matrix renormalization group (TMRG) algorithm \cite{TMRG} to evaluate partion functions in 2D classical systems. However, iTEBD differs from them at its core in two important aspects: first, TMRG, PWFRG and iDMRG are variational methods, while iTEBD amounts to a power method; second, whereas TMRG, PWFRG and iDMRG converge towards an infinite system by adding sites to a finite lattice, in iTEBD the system is infinite from the onset. We notice that Ref. \cite{iDMRG}, which contains a useful comparative study, highlights that the iTEBD is significantly more accurate than other proposals in determining ground states. Nevertheless, all these methods are of comparable interest.
 
The rest of the paper is organized as follows. Sect. II explains how to obtain the canonical form of an iMPS by orthonormalizing all its bond indices. Sect. III presents the generalization the iTEBD algorithm to account for non-unitary evolution. Sect. IV discusses an application of the algorithm to 2D classical lattice models and Sect. V contains some conclusions. Finally, the appendix presents a detailed description of how to implement the algorithm for very specific forms of the evolution operator.

\section{Canonical form of an infinite MPS}

\begin{figure}[h]
\includegraphics[width=0.5\textwidth]{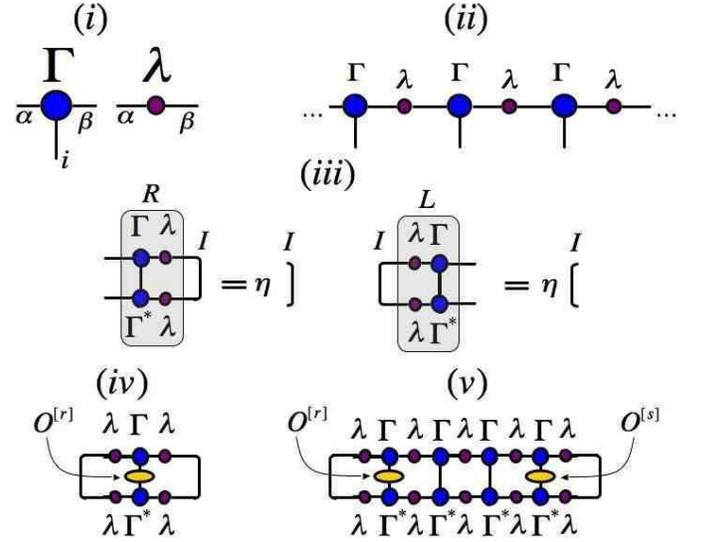}
\caption{(color online) $(i)$ Diagrammatic representation of the tensors $\Gamma$ and $\lambda$ that form an iMPS for a pure state $\ket{\Psi}$ of a translationally invariant chain. ($ii$) The iMPS is actually an infinite one-dimensional tensor network consisting of alternating copies of the tensors $\Gamma$ and $\lambda$. In the diagram, a leg shared by two tensors corresponds to an index over which there is an implicit sum. ($iii$) Diagrammatic representation of the conditions in Eqs. (\ref{canonicalR})-(\ref{canonicalL}), which are fulfilled if and only if the iMPS $\{ \Gamma, \lambda \}$ is in its canonical form. ($iv$) Tensor network corresponding to the expectation value $\bra{\Psi}O^{[r]}\ket{\Psi}$ of an operator $O^{[r]}$, acting on site $r$, when $\ket{\Psi}$ is represented by an iMPS $\{\Gamma, \lambda \}$ in its canonical form. As a result of the orthonormality of the Schmidt vectors in Eq. (\ref{Schmidt}), only a very small number of tensors need to be considered. ($v$) Tensor network for the computation of the two-point correlator $\bra{\Psi}O^{[r]} O^{[s]}\ket{\Psi}$ when $\ket{\Psi}$ is represented by an iMPS $\{\Gamma,\lambda\}$ in the canonical form.}
\label{fig1}
\end{figure}

We consider an infinite 1D lattice, where each site is labeled by an integer $r \in \mathbb{Z}$ and described by a Hilbert space $\mathbb{C}^{d}$ of finite dimension $d$. The lattice is in a pure state $\ket{\Psi}\in \bigotimes_{r\in\mathbb{Z}} \mathbb{C}^{d}$ that is invariant under translations by $n$ sites (and multiples thereof). Following Ref. \cite{iTEBD}, we represent $\ket{\Psi}$ using an iMPS, which in the simplest case ($n=1$) consists of a pair of tensors $\{\Gamma, \lambda\}$, see Figs. (\ref{fig1}.$i$)-(\ref{fig1}.$ii$).  Here $\Gamma$ is made of complex coefficients $\Gamma_{\alpha\beta}^{i}$, with two \emph{bond} indices $\alpha$ and $\beta$ $(\alpha,\beta = 1,\cdots,\chi)$ and one \emph{physical} index $i$ ($i=1,\cdots,d$) that labels an orthonormal basis in $\mathbb{C}^{d}$, whereas $\lambda$ is a diagonal matrix with non-negative diagonal elements $\lambda_{\alpha}$. The integer $\chi$ is known as the rank of the iMPS.

We say that an iMPS $\{\Gamma,\lambda\}$ is in its \emph{canonical form} \cite{TEBD,tTN} when, on each bond, index $\alpha$ is related to the Schmidt decomposition of $\ket{\Psi}$,
\begin{equation}
 \ket{\Psi} = \sum_{\alpha=1}^{\chi} \lambda_{\alpha} \ket{\Phi^{L}_{\alpha}}\otimes\ket{\Phi^{R}_{\alpha}} \ ,
\label{Schmidt}
\end{equation} 
that is, when the diagonal matrix $\lambda$ contains the decreasingly ordered Schmidt coefficients ($\lambda_1 \geq \lambda_2 \geq \cdots \geq \lambda_{\chi} \geq 0$) and $\alpha$ labels the Schmidt vectors, which form orthonormal sets, $\braket{\Phi^{L}_{\alpha}}{\Phi^{L}_{\alpha'}} =\braket{\Phi^{R}_{\alpha}}{\Phi^{R}_{\alpha'}} = \delta_{\alpha\alpha'}$. In terms of the matrices $R$ and $L$, defined as
\beqa
R_{(\alpha \alpha'),(\beta \beta')} &\equiv& \sum_{i = 1}^d \left( \Gamma^{i}_{\alpha \beta} \lambda_\beta \right) \left( \Gamma^{i}_{\alpha' \beta'} \lambda_{\beta'}\right)^* \label{transferR}\\
L_{(\alpha \alpha'),(\beta \beta')} &\equiv& \sum_{i = 1}^d \left( \lambda_\alpha \Gamma^{i}_{\alpha \beta} \right) \left( \lambda_{\alpha'} \Gamma^{i}_{\alpha' \beta'} \right)^* \,
\label{transferL}
\eeqa
the canonical form corresponds to the conditions
\beqa
\sum_{\beta,\beta'} R_{(\alpha \alpha'),(\beta \beta')} \delta_{\beta\beta'} &=& \eta \delta_{\alpha\alpha'} \label{canonicalR} \\
\sum_{\alpha,\alpha'} \delta_{\alpha\alpha'} L_{(\alpha \alpha'),(\beta \beta')}  &=& \eta \delta_{\beta\beta'} \ ,
\label{canonicalL}
\eeqa
where $\eta \in \mathbb{C}$. In other words, the identity operator $I_{\alpha \alpha'} = \delta_{\alpha \alpha'}$ is a right (left) eigenvector of matrix $R$ (respectively $L$) with eigenvalue $\eta$, see Fig.(\ref{fig1}.$iii$). Incidentally, $\eta$ is the dominant eigenvalue \cite{dominant} of both $R$ and $L$, and is equal to 1 if and only if $\ket{\Psi}$ is normalized \cite{unique}. 

Two good reasons to express an iMPS in its canonical form are the following. First, it facilitates the computation of expectation values for local operators. For instance, for $O^{[r]}$ an operator acting on site $r$, orthogonality of the Schmidt bases implies that $\bra{\Psi}O^{[r]}\ket{\Psi}$ is simply
\begin{equation}
	\bra{\Psi}O^{[r]}\ket{\Psi} = \sum_{\alpha,i,\beta} (\lambda_{\alpha})^2\Gamma_{\alpha\beta}^{i} O^{[r]}_{ij} (\Gamma_{\alpha\beta}^{j})^* (\lambda_{\alpha})^2 \ ,
\end{equation}
which can be computed in $O(d^2 \chi^2)$ time. 
Similarly, the expression for a two-point correlator $\bra{\Psi}O^{[r]}O^{[s]}\ket{\Psi}$ involves only of the order of $|s-r|$ tensors, see Figs. (\ref{fig1}.$iv$)-(\ref{fig1}.$v$). Second, as we will discuss in Sect. III, the canonical form simplifies the truncation of bond indices, a process that is necessary in order to prevent the rank of the iMPS from growing during a simulation.

\begin{figure}[h]
\includegraphics[width=8cm]{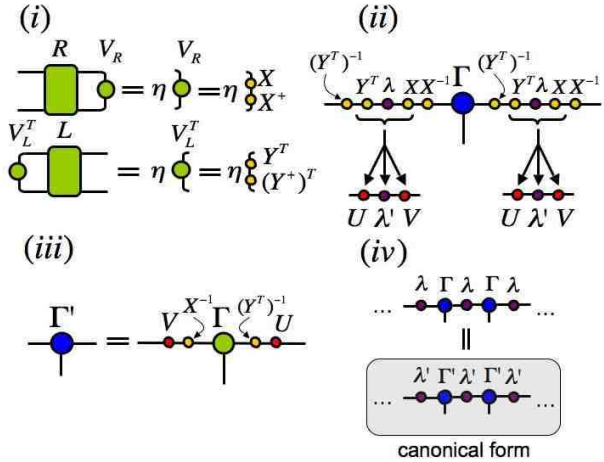}
\caption{(color on-line) ($i$-$iii$) The three steps involved in the computation of the canonical form $\{\Gamma',\lambda'\}$ for an iMPS $\{\Gamma,\lambda\}$ for $\ket{\Psi}$, as explained in the text. In particular, $\lambda'$ is obtained in step ($ii$) as the singular values of $Y^T\lambda X$,  whereas $\Gamma'$ is defined in step ($iii$). $(iv)$ Both $\{\Gamma,\lambda\}$ and $\{\Gamma',\lambda'\}$ represent the same state $\ket{\Psi}$, as can be verified by direct substitution.}
\label{fig2}
\end{figure}

Theorem 1 of Ref. \cite{tTN} explains how to bring an MPS to its canonical form in the case of a finite chain. This is done by orthonormalizing bond indices, starting from one boundary of the chain and progressing through the whole system, with a cost proportional to its length. In the infinite case, we use translational invariance to reduce this cost to a constant. More specifically, given an iMPS $\{\Gamma,\lambda\}$ for $\ket{\Psi}$, we can obtain a canonical form $\{\Gamma', \lambda'\}$ for the same state through three steps, illustrated in Fig. (\ref{fig2}):

$(i)$ Find the matrix $V_R$ that is the dominant right eigenvector \cite{dominant} of $R$, in the sense of Fig. (\ref{fig2}.$i$), with dominant eigenvalue $\eta \in \mathbb{C}$ (here $\eta$ is assumed to be unique \cite{eta}). Similarly, find the matrix $V_L$ that is the dominant left eigenvector of $L$, which also has eigenvalue $\eta$ [we use a large-scale, non-Hermitian eigenvalue solver \cite{Shi03}, such as an Arnoldi method, and exploit the tensor network structure of $R$ and $L$].
Decompose matrices $V_R$ and $V_L$, which are Hermitian and non-negative (since they originate in the scalar product of a set of non-orthogonal vectors), as squares $V_R = XX^{\dagger}$ and $V_L = Y^{\dagger}Y$. For instance, if $V_L = W D W^{\dagger}$ is the eigenvalue decomposition of $V_L$, then $Y^{\dagger} = W \sqrt{D}$ and $Y = \sqrt{D} W^{\dagger}$ \cite{Cho}.

$(ii)$ Introduce the two resolutions of the identity matrix $\mathbb{I} = (Y^T)^{-1}Y^T$ and $\mathbb{I} = X X^{-1}$ in the bond indices of the iMPS as indicated in Fig.(\ref{fig2}.$ii$). Then, compute the singular value decomposition of the product $Y^T \lambda X$, namely $Y^T \lambda X = U \lambda' V$, where $U,V$ are unitary and the diagonal matrix $\lambda'$ contains the Schmidt coefficients of $\ket{\Psi}$.

$(iii)$ Arrange the remaining tensors $V$, $X^{-1}$, $\Gamma$,  $(Y^T)^{-1}$ and $U$ into a new tensor $\Gamma'$ as in Fig. (\ref{fig2}.$iii$).

\begin{figure}[h]
\includegraphics[width=8cm]{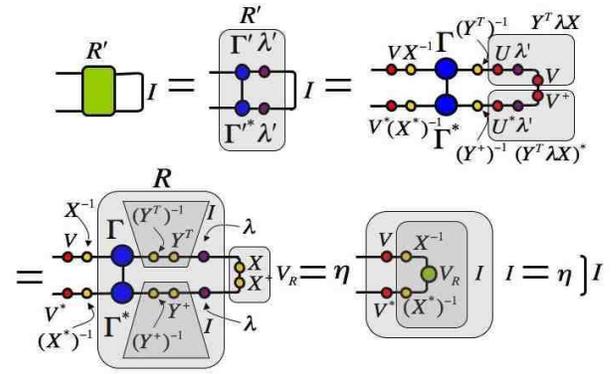}
\caption{(color online) Proof that $\{\Gamma'\lambda'\}$, the iMPS obtained from $\{\Gamma,\lambda\}$ by re-orthonormalizing its bond indices following Fig. (\ref{fig2}), is indeed in the canonical form. The diagram only shows the condition of Eq. (\ref{canonicalR}) for $R'$. The proof of Eq. (\ref{canonicalL}) for $L'$ is analogous.}
\label{fig3}
\end{figure}

All the previous manipulations can be implemented with computational cost scaling as $O(d\chi^3)$. A proof that the resulting iMPS $\{\Gamma',\lambda'\}$ is indeed in the canonical form is given in Fig. (\ref{fig3}). 

\begin{figure}[h]
\includegraphics[width=8cm]{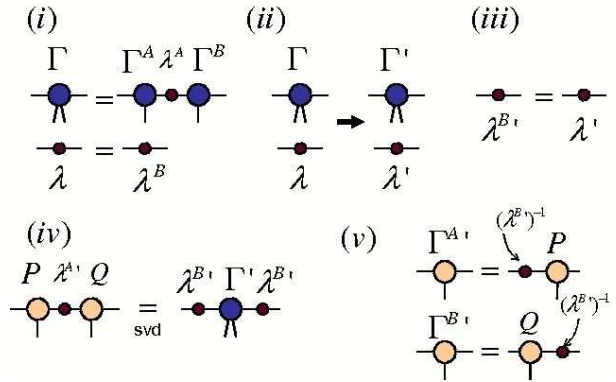}
\caption{(color online) Construction of the canonical form $\{{\Gamma^{A}}'$, ${\lambda^{A}}'$, ${\Gamma^{B}}'$, ${\lambda^{B}}'\}$ from an iMPS $\{\Gamma^{A}$, $\lambda^{A}$, $\Gamma^{B}$, $\lambda^{B}\}$ that is invariant under translations by $n=2$ sites: ($i$) Coarse-grained iMPS $\{\Gamma,\lambda\}$. ($ii$) Canonical form $\{\Gamma',\lambda'\}$ for the coarse-grained iMPS$\{\Gamma,\lambda\}$. ($iii$) ${\lambda^B}'$ corresponds to $\lambda'$. ($iv$) ${\lambda^A}'$ is obtained through a singular value decomposition, from where also ($v$) ${\Gamma^A}'$ and ${\Gamma^B}'$ are obtained after minor manipulations.}
\label{figExtra}
\end{figure}

We can now analyze the case where $\ket{\Psi}$ is invariant under translations by $n>1$ sites. 
For $n=2$, the state $\ket{\Psi}$ is represented by an iMPS that consists of four alternating tensors $\{\Gamma^{A}$, $\lambda^{A}$, $\Gamma^{B}$, $\lambda^{B}\}$, where $A$ and $B$ denote odd and even sites in the chain \cite{iTEBD}. The canonical form $\{{\Gamma^{A}}'$, ${\lambda^{A}}'$, ${\Gamma^{B}}'$, ${\lambda^{B}}'\}$, defined as before to correspond to the Schmidt decomposition at each bond, can be obtained as follows. First we coarse-grain the chain by regarding each pair of sites $AB$ as a single site and represent $\ket{\Psi}$ with an iMPS $(\Gamma,\lambda)$ as in the $n=1$ case. Then we transform the coarse-grained iMPS $(\Gamma,\lambda)$ into its canonical form $(\Gamma',\lambda')$. Finally, we split $\Gamma'$ into three tensors ${\Gamma^{A}}'$, ${\lambda^{A}}'$ and ${\Gamma^{B}}'$ by means of a singular value decomposition, see Fig (\ref{figExtra}). These steps can be implemented with a computational cost that scales as $O(d^3 \chi^3)$. The case of a generic $n$ is addressed similarly, and the computational cost scales as $O(nd^3 \chi^3)$.

\section{Simulation of non-unitary evolution}

\begin{figure}[h]
\includegraphics[width=8cm]{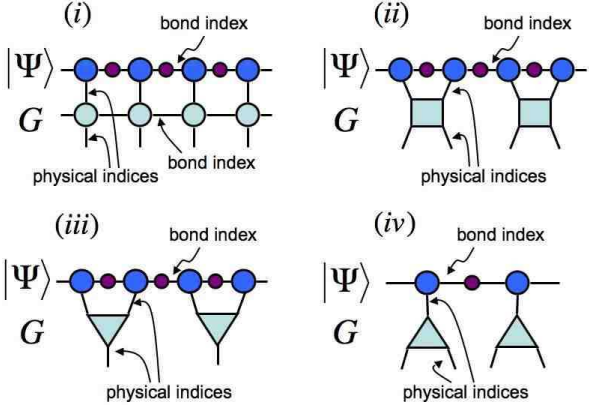}
\caption{(color online) Four types of gates $G$ acting on an iMPS: $(i)$ infinite matrix product operator (iMPO); $(ii)$ product of two-site operators; $(iii)$ product of two-to-one-site operators; $(iv)$ Tensor product of one-to-two-sites operators. Detailed explanations on how to update the iMPS for cases ($ii$)-($iv$) can be found in the appendix. Notice that gates ($iii$) and ($iv$) change the number of sites in the lattice. These gates appear e.g. when coarse-graining (or fine-graining) the chain.}
\label{TEBD1}
\end{figure}

In this section we discuss how to update the iMPS for state $\ket{\Psi}$ after a gate $G$ acts on the entire lattice. That is, we aim to build an iMPS for the resulting state $\ket{\Psi'} = G\ket{\Psi}$. We assume that $G$ is expressed as a one-dimensional tensor network (of some sort) that is invariant under translations by $n$ sites, see Fig. (\ref{TEBD1}) for several examples. As a remark, let us mention that non-unitary gates such as the ones in Fig. (\ref{TEBD1}) appear in the so-called iPEPS algorithm to simulate 2D quantum lattice systems \cite{iPEPS}. 

\begin{figure}[h]
\includegraphics[width=7cm]{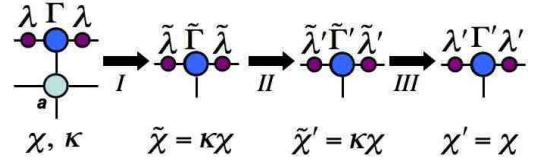}
\caption{(color online) Sequence of transformations (I)-(III) that produce a truncated iMPS $\{\Gamma'\lambda'\}$ for $\ket{\Psi'}=G\ket{\Psi}$ from an iMPS $\{\Gamma,\lambda\}$ for $\ket{\Psi}$, as explained in the text. Notice in particular that, as a result of the truncation step, the initial and final iMPSs have the same rank $\chi$.}
\label{fig4}
\end{figure}

We focus again on the case $n=1$  and, for concreteness, we assume $G$ is specified by an infinite \emph{matrix product operator} (iMPO) as in Fig. (\ref{TEBD1}.$i$). This iMPO is represented by a tensor $a$ of complex components $a_{\mu\nu}^{ij}$, where $i$ and $j$ are physical indices and $\mu$ and $\nu$ ($\mu,\nu= 1,\cdots, \kappa$) are bond indices. The update occurs in three steps, illustrated in Fig. (\ref{fig4}): 

(I) \emph{Contraction:} the tensors $\{\Gamma,\lambda\}$ for $\ket{\Psi}$ are contracted with the tensors that specify the gate $G$, producing an iMPS $\{\tilde{\Gamma},\tilde{\lambda}\}$ for $\ket{\Psi'}$,
\begin{equation}
	\tilde{\Gamma}_{\tilde{\alpha}\tilde{\beta}}^{j} \equiv \sum_{i=1}^d \Gamma_{\alpha\beta}^{i} a^{ij}_{\mu\nu}, ~~~~~~~~~~~
	\tilde{\lambda}_{\tilde{\beta}} \equiv \lambda_{\beta}~~~~~(\forall \nu).
\end{equation}
Here indices $\tilde{\alpha}$ and $\tilde{\beta}$ ($\tilde{\alpha},\tilde{\beta} = 1,\cdots, \tilde{\chi}$) are defined as $\tilde{\alpha} \equiv (\alpha,\mu)$ and $\tilde{\beta} \equiv (\beta,\nu)$. Notice that the rank $\tilde{\chi}$ of the new iMPS,  $\tilde{\chi} \equiv \kappa\chi$, is larger than the rank $\chi$ of the initial iMPS. The computational cost of this step is $O(d^2 \kappa^2 \chi^2)$. 

(II) \emph{Orthogonalization:} the iMPS $\{\tilde{\Gamma},\tilde{\lambda}\}$ for $\ket{\Psi'}$ is brought into its canonical form $\{{\tilde{\Gamma}}',$ ${\tilde{\lambda}}'\}$ with a cost that scales as $O(d \kappa^3 \chi^3)$.

(III) \emph{Truncation:} a final iMPS $\{\Gamma',$ $\lambda'\}$ is obtained from the canonical form $\{{\tilde{\Gamma}}',$ ${\tilde{\lambda}}'\}$ by truncating all bond indices. In particular, on each bond we preserve the first $\chi$ values of the index, corresponding to the $\chi$ largest Schmidt coefficients.

The net result is an \emph{approximate} iMPS $\{\Gamma',\lambda'\}$ for $\ket{\Psi'}$, obtained with a total computational cost $O(d^2 \kappa^2 \chi^2 + d\kappa^3\chi^3)$ \cite{bound}. The truncation step is necessary in order to keep the rank $\chi$ (and therefore the computational cost) constant during a simulation, where typically not just one gate $G$ but rather of a whole series $\{G_1,G_2,\cdots\}$ is sequentially applied to the chain. The truncation of the bond indices introduces an error that is hard to evaluate in an infinite system. Here we apply, to all bond indices simultaneously, the truncation scheme that is known to be optimal when applied only to one bond index \cite{truncation}. 

For $n>1$, as is the case e.g. in Fig. (\ref{TEBD1}.$ii$), we can again coarse-grain the system and proceed as in the $n=1$ case, which will result in a iMPS $\{\Gamma',\lambda'\}$. Then $\Gamma'$ is broken into several other tensors $\{{\Gamma^{A}}', {\lambda^{A}}', {\Gamma^{B}}', \cdots \}$, process that may require additional truncations. See the appendix for a detailed analysis of some particular cases.

We are therefore able to address non-unitary evolution on an infinite chain. When the gate $G$ breaks into a row of two-site gates as in Fig. (\ref{TEBD1}.ii), and each two-site gate is unitary, 
then the canonical form of the iMPS is preserved (up to truncation errors) without need of the orthogonalization step, recovering the original formulation of the iTEBD algorithm \cite{iTEBD}. Notice that in Ref. \cite{iTEBD} the algorithm is also used to compute the ground state of the system. This is done by simulating (non-unitary) imaginary time evolution 
\begin{equation}
	\ket{\Psi_{\tau}} = \frac{\exp(-H\tau)\ket{\Psi_0}}{||\exp(-H\tau)\ket{\Psi_0}||},
\end{equation}
where $H$ is the Hamiltonian of the infinite chain and $\ket{\Psi_0}$ some initial state, and by exploiting the fact that under proper circumstances the ground state $\ket{\Psi}$ of $H$ is the fixed point of such evolution,
\begin{equation}
	\ket{\Psi} = \lim_{\tau\rightarrow \infty }\frac{\exp(-H\tau)\ket{\Psi_0}}{||\exp(-H\tau)\ket{\Psi_0}||}.
\end{equation}
We emphasize that such calculation succeeds thanks to a fortunate combination of favorable, unlikely circumstances. When the simulation is performed using small time steps, two-site gates that are close to the identity operator are used. These gates destroy the canonical form of an initial iMPS, but they leave its bond indices in a (non-orthonormal) basis that still seems to lead to reasonably small errors during their truncation. One would expect the bond bases to become less and less adequate for truncation over time, as the accumulated $\tau$ increases, since the overall evolution $\exp(-H\tau)$ departs more and more from the identity. But it turns out that the singular value decomposition used in order to update the iMPS at each time step has the effect of reorganizing the indices favorably, partially compensating the non-unitary effects \cite{compensation}. Finally, all the excessive truncation errors introduced during the simulation are washed away at its final stages, where increasingly small time steps are used. These have the intended effect of reducing Suzuki-Trotter errors \cite{TEBD}, but they also imply that the gates become almost unitary (that is, very close to the identity). One can see that, as a result, by the end of the simulation the iMPS approximation for the ground state is not only accurate, but it is also very close to the canonical form. 

Thanks to actively transforming the iMPS into its canonical form, the present extension is not restricted to unitary evolution and can be applied to a wider range of 1D problems. In particular, it can be used to simulate master equation evolution and to compute thermal states using the mixed state formalism of Refs. \cite{ZV}. Importantly, it can also be used to manipulate the state of an infinite 2D lattice, both for classical (see example below) and quantum systems \cite{iPEPS}. This is achieved after the 2D problem is recast into that of finding the dominant eigenvalue $\theta$ and dominant eigenvector \cite{dominant} $\ket{\Psi}$ of a 1D transfer matrix $T$ that decomposes into a finite sequence $\{G_1,G_2, \cdots, G_m\}$ of gates. The dominant eigenvector satisfies
\begin{equation}
	\ket{\Psi} = \lim_{p \rightarrow \infty} \frac{T^{p}\ket{\Psi_0}}{||T^{p}\ket{\Psi_0}||},
	\label{eq:dominant}
\end{equation}
and is obtained by simulating the repeated application of $T$ on an initial state $\ket{\Psi_0}$, until converge is attained. The dominant eigenvalue can be obtained from the dominant eigenvector $\ket{\Psi}$ and any other vector $\ket{\Phi}$, $\braket{\Phi}{\Psi} \neq 0$, since
\begin{equation}
	\theta = \frac{\bra{\Phi}T\ket{\Psi}}{\braket{\Phi}{\Psi}}.
\end{equation}
In the next section we provide an explicit example of calculation of dominant eigenvalue and eigenvector of a one-dimensional transfer matrix.
 
\section{Example: 2D classical systems}
 
\begin{figure}[h]
\includegraphics[width=8cm]{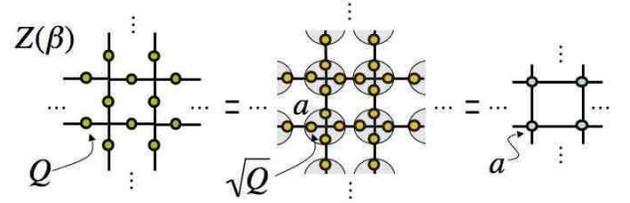}
\caption{(color online) The partition function $Z(\beta)$ of a 2D classical system can be written as the contraction of a 2D tensor network. In the case of an infinite square lattice with isotropic and homogeneous interactions, this tensor network consists of infinitely many copies of the tensor $a$ in Eq. (\ref{eq:a}).}
\label{partition}
\end{figure}

In this section we explain how the above algorithm can be used to compute the partition function, local observables and two-point correlators of a classical spin system. We consider an infinite 2D lattice where each site, labeled by a vector $\vec{r}$, contains a $d$-dimensional spin $s^{[\vec{r}]}$ that interacts with nearest neighbor spins according to a Hamiltonian $K$,
\begin{equation}
	K(\{s\}) = \sum_{\langle \vec{r},\vec{r}' \rangle} K_2(s^{[\vec{r}]},s^{[\vec{r}']}).
\end{equation}
The system's partition function reads
\begin{equation}
Z(\beta) = \sum_{\{ s \}} e^{-\beta K(\{ s \})} = \sum_{\{ s \}} \prod_{\langle \vec{r},\vec{r}' \rangle} e^{-\beta K_2(s^{[\vec{r}]},s^{[\vec{r}']})}, 
\label{eq:partition}
\end{equation}
where $\beta$ is the inverse temperature. For concreteness, we consider a square lattice with an isotropic interaction $K_2$. Let $\sqrt{Q}$ denote the squared root of the Hermitian matrix $Q_{ss'}\equiv \exp(-\beta K_2(s,s'))$ \cite{no_square_root}. We can express the partition function as the contraction of an infinite 2D tensor network specified by a single tensor $a$, 
\begin{equation}
	a_{ijkl} \equiv \sum_{s} (\sqrt{Q})_{i s} (\sqrt{Q})_{j s} (\sqrt{Q})_{q s} (\sqrt{Q})_{l s}
\label{eq:a}
\end{equation}
that is repeated on all sites \cite{PEPSclas}, see Fig. (\ref{partition}). The above tensor can be computed in $O(d^5)$ time.

\begin{figure}[h]
\includegraphics[width=8cm]{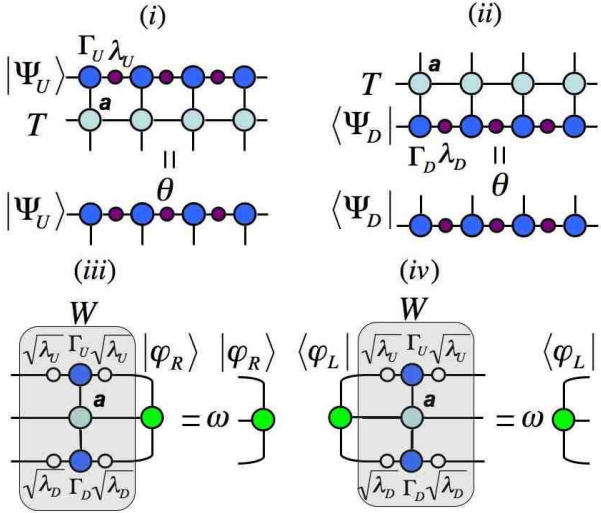}
\caption{(color online) Computation of the partition function through Eq. (\ref{eq:ZOmega}). ($i$) $\ket{\Psi_U}$ is the up dominant eigenvector of the transfer matrix $T$, with dominant eigenvalue $\theta$. ($ii$) Similarly, $\ket{\Psi_D}$ is a down dominant eigenvector of $T$, with dominant eigenvalue $\theta$. ($iii$) $\ket{\varphi_R}$ is the right dominant eigenvector of matrix $W$, with dominant eigenvalue $\omega$. ($iv$) $\ket{\varphi_L}$ is the left dominant eigenvector of matrix $W$, with dominant eigenvalue $\omega$.}
\label{fig:ThetaOmega}
\end{figure}

We now introduce an infinite 1D transfer matrix $T$ consisting of one row of tensors $a$, see Fig. (\ref{fig:ThetaOmega}). Then we have
\begin{equation}
	Z(\beta) = \lim_{p\rightarrow \infty} \tr (T^p) = \lim_{p\rightarrow \infty} \theta^{p},
\end{equation}
where $\theta$ is the dominant eigenvalue of $T$. Let $\ket{\Psi_U}$ and $\ket{\Psi_D}$ be the corresponding (up and down) eigenvectors,
\begin{equation}
	T\ket{\Psi_U} = \theta \ket{\Psi_U},~~~~~~~~~~~\bra{\Psi_D} T = \theta \bra{\Psi_D},
\end{equation}
that we normalize to $\braket{\Psi_D}{\Psi_U}=1$. Then
\begin{equation}
	\theta = \bra{\Psi_D}T\ket{\Psi_U} = \tr(W^q) = \lim_{q \rightarrow \infty} \omega^q,
\end{equation}
where $\omega$ is the dominant eigenvalue of matrix $W$ defined in Fig. (\ref{fig:ThetaOmega}), and we finally have
\begin{equation}
	Z(\beta) = \lim_{p,q\rightarrow \infty} \omega^{p ~q}.
\label{eq:ZOmega}
\end{equation}

Therefore, in order to evaluate the partition function $Z(\beta)$, we will first construct an iMPS $\{\Gamma^U, \lambda^U\}$ for $\ket{\Psi_U}$ and an iMPS $\{\Gamma^D, \lambda^D\}$ for $\ket{\Psi_D}$ by iteratively applying the transfer matrix $T$ on an initial state $\ket{\Psi_0}$ (c.f. Eq. (\ref{eq:dominant})). Specifically, we use the iTEBD algorithm as discussed in the previous section to simulate the state 
\begin{equation}
	\ket{\Psi_{p}} \equiv \frac{T^{p}\ket{\Psi_0}}{||T^{p}\ket{\Psi_0}||},~~~~~~~~p=1,2,\cdots
\label{eq:Psip}
\end{equation}
for increasing values of $p$, until the resulting iMPS has converged within some agreed precision \cite{converged}. The computation is approximate, in that an iMPS with finite rank $\chi$ will be used to represent the dominant eigenvectors, which in general may only be represented exactly with an infinite rank $\chi$. Notice that if $K_2$ is isotropic, the transfer matrix can be made Hermitian, in which case $\bra{\Psi_D} = \ket{\Psi_U^{\dagger}}$ [otherwise $\ket{\Psi_D}$ also needs to be computed]. 
From the converged iMPSs $\{\Gamma^U, \lambda^U\}$ and $\{\Gamma^D, \lambda^D\}$ we can construct matrix $W$. The dominant eigenvalue $\omega$ of $W$ can then be computed using a large-scale eigenvalue solver and exploiting its tensor network structure in $O(d^2 \chi^3 + d^4 \chi^2)$ time. 
 
\begin{figure}[h]
\includegraphics[width=8cm]{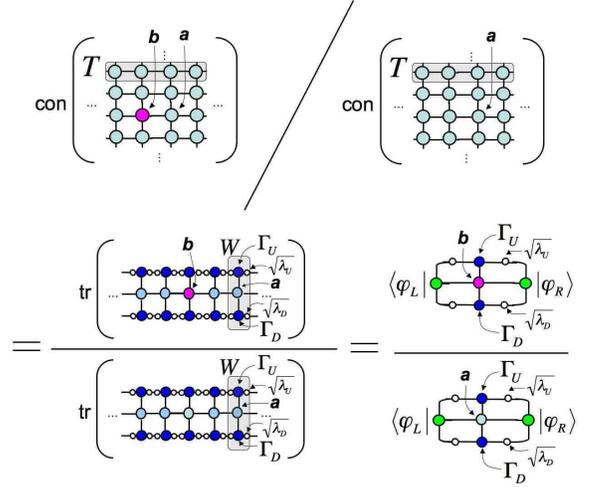}
\caption{(color online) The expectation value $\langle f(s^{[\vec{r}]})\rangle$ from Eq. (\ref{eq:classobs}) is the ratio of the contraction of two infinite 2D tensor networks. By introducing the dominant eigenvectors of the one-dimensional transfer matrix $T$, we can rewrite $\langle f(s^{[\vec{r}]})\rangle$ as the ratio of the trace of two infinite 1D tensor networks. Finally, by introducing the dominant eigenvectors of matrix $W$, we obtain a ration of two simple tensor networks.}
\label{fig:LocalObservable}
\end{figure}

On the other hand, for any function $f(s)$ of one spin, the expectation value
\begin{equation}
\langle f(s^{[\vec{r}]})\rangle = \frac{1}{Z(\beta)}\sum_{\{s\}} f(s^{[\vec{r}]}) e^{-\beta K(\{s\})} \ ,
\label{eq:classobs}
\end{equation}
is, up to the factor $1/Z(\beta)$, also given by the contraction of an infinite 2D tensor network, obtained from that for $Z(\beta)$ by replacing tensor $a$ on site $\vec{r}$ with tensor $b$,
\begin{equation}
	b_{ijkl} \equiv \sum_{s} f(s) (\sqrt{Q})_{i s} (\sqrt{Q})_{j s} (\sqrt{Q})_{q s} (\sqrt{Q})_{l s} \ ,
\end{equation}
again computable in $O(d^5)$ time. As illustrated in Fig. (\ref{fig:LocalObservable}), $\langle f(s^{[\vec{r}]})\rangle$ is eventually written as the ratio of two small tensor networks. These tensor networks are expressed entirely in terms of: tensors $a$ and $b$; tensors $\{\Gamma^U,\lambda^U\}$ and $\{\Gamma^D\lambda^D\}$ defining the dominant eigenvectors $\ket{\Psi_U}$ and $\ket{\Psi_D}$ of the one-dimensional transfer matrix $T$; and the dominant vectors $\ket{\varphi_R}$ and $\ket{\varphi_L}$ of the matrix $W$. We have already indicated how to proceed in the computation of these quantities. 

\begin{figure}[h]
\includegraphics[width=6cm]{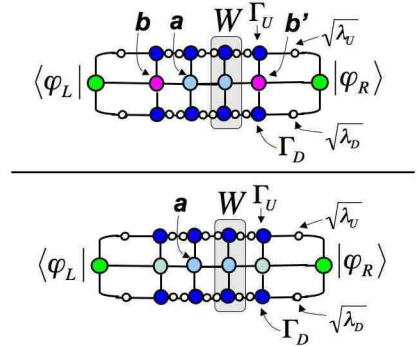}
\caption{(color online) Following steps analogous as those of Fig. (\ref{fig:LocalObservable}) for the expectation value $\langle f(s^{[\vec{r}]})\rangle$, the two-point correlator $\langle f(s^{[\vec{r}]})g(s^{[\vec{r}']})\rangle$ can also be reduced to the ratio of two simple tensor networks.}
\label{fig:TwoPoint}
\end{figure}

Similarly, we can build a tensor network for the expectation value of the correlator
\begin{equation}
\langle f(s^{[\vec{r}]})g(s^{[\vec{r}']})\rangle = \frac{1}{Z(\beta)}\sum_{\{s\}} f(s^{[\vec{r}]})g(s^{[\vec{r}']}) e^{-\beta K(\{s\})} \ ,
\label{eq:classcorr} 
\end{equation}
by replacing the tensor $a$ in sites $\vec{r}$ and $\vec{r}'$ with appropriate tensors $b$ and $b'$ and proceeding in a similar way as the previous case, see Fig. (\ref{fig:TwoPoint}). Notice that we assume that sites $\vec{r}$ and $\vec{r}'$ lie on the same row of the lattice.

\begin{figure}
\includegraphics[width=0.5\textwidth]{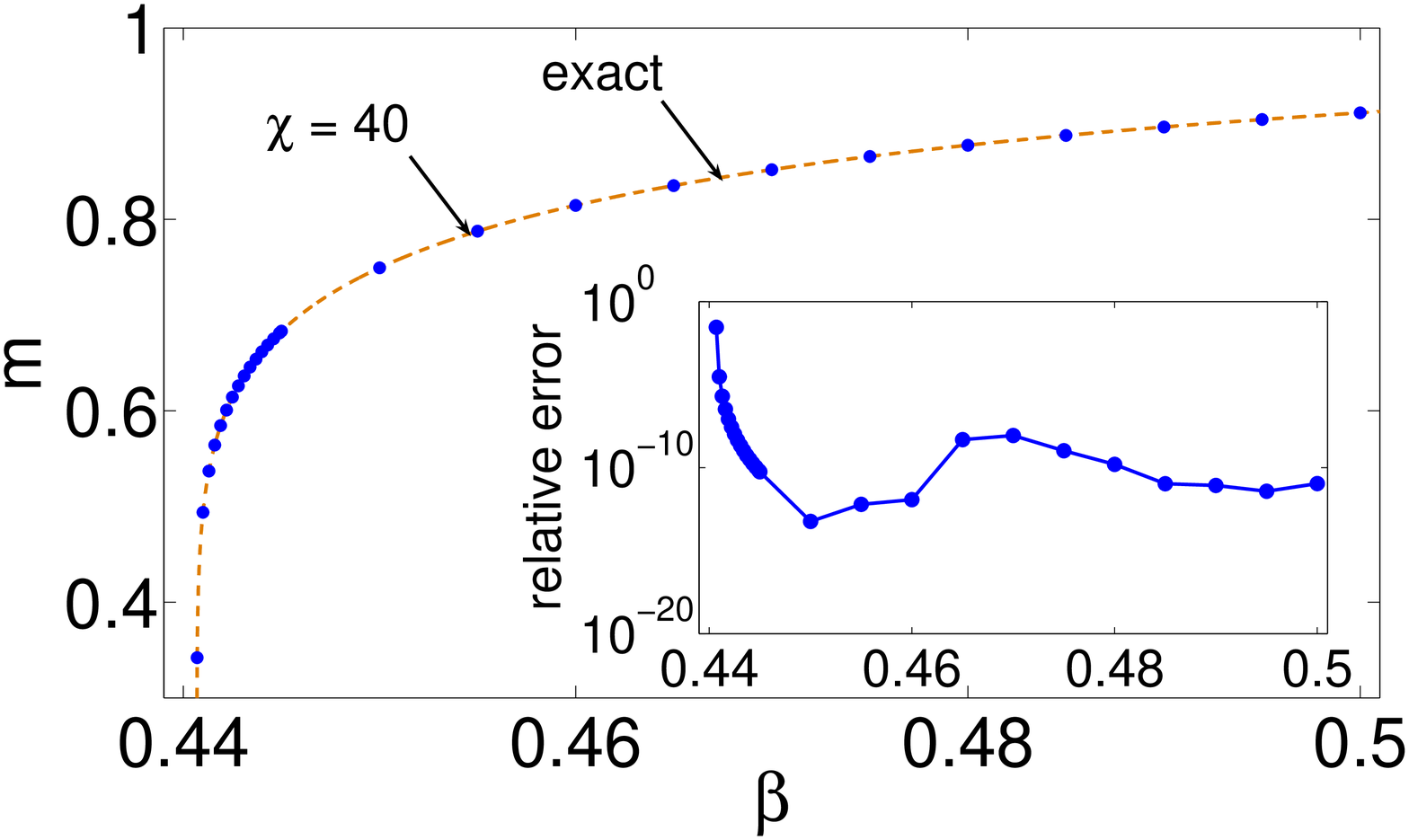}
\caption{(color online) Magnetization per lattice site for the infinite-size 2D classical Ising model at different temperatures $\beta$. The exact solution $m = (1-((\sinh{(2\beta)})^{-4}))^{1/8}$ has been included. The numerical results have been obtained by approximating the dominant eigenvectors of the one-dimensional transfer matrix $T$ with an iMPS of rank $\chi=40$. The inset shows the relative error.}
\label{fig:IsingMag}
\end{figure}

\begin{figure}
\includegraphics[width=0.52\textwidth]{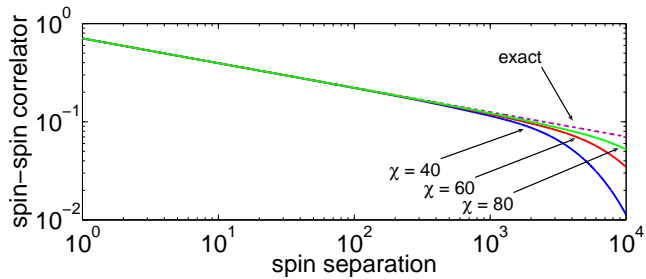}
\caption{(color online) Two-point correlators for the infinite-size 2D classical Ising model at critical temperature, $\beta_c = \frac{1}{2}\log(1+\sqrt{2})$, along a row of the square lattice. The exact solution, which scales as $\langle s^{[\vec{r}]}s^{[\vec{r'}]}\rangle_{\beta_c} \approx c|\vec{r}-\vec{r}'|^{-1/4}$, has been included. The numerical results have been obtained by approximating the dominant eigenvectors of the one-dimensional transfer matrix $T$ with an iMPS of rank $\chi=40$, $\chi=60$ and $\chi=80$.}
\label{fig:IsingCorr}
\end{figure}

Fig. (\ref{fig:IsingMag}) shows the magnetization per site $m \equiv \langle s^{[\vec{r}]}\rangle$ for the 2D Ising model, defined by 
\begin{equation}
	K_2(s,s') = -s~ s'  ~~~~~~~~~~~~ s,s' = \pm 1 \ ,
\end{equation}
at different values of $\beta$. We have used an iMPS of rank $\chi=40$ to represent the dominant eigenvectors $\ket{\Psi_U}$ and $\ket{\Psi_D}$, and proceeded as explained above. It is noteworthy that the numerical results reproduce the exact behaviour of $m$ with small relative error. Furthermore, Fig. (\ref{fig:IsingCorr}) shows the decay with $|\vec{r}-\vec{r}'|$ of the spin-spin correlator $\langle s^{[\vec{r}]}s^{[\vec{r}']}\rangle$ at the critical point, $\beta_c = \frac{1}{2}\log(1+\sqrt{2})$. In this case we have used an iMPS of rank $\chi=40,60,80$. Remarkably, the numerical results reproduce the correct power-law decay $\sim |\vec{r}-\vec{r}'|^{-1/4}$ for distances up to thousands of spins, with increasing accuracy as $\chi$ increases.

\section{Conclusions}

In this paper we have explained how to extend the iTEBD algorithm so that it can be applied to simulate any evolution that can be expressed as a sequence of one-dimensional tensor networks. The key new ingredient is a recipe to rewrite any iMPS in the canonical form, as required in order to properly truncate the bond indices.

The iTEBD algorithm can therefore be applied to simulate not only unitary evolution, but also master equation evolution and imaginary time evolution. It can also be used to find the dominant eigenvalue and dominant eigenvector of a one-dimensional transfer matrix. This last application is particularly relevant in order to analyze the partition function of 2D classical models, as we explained, and it also plays a prominent role in the iPEPS algorithm for 2D quantum systems \cite{iPEPS} and some of the MERA algorithms \cite{MERA}.

{\it Acknowledgements.---} We acknowledge discussions with Jacob Jordan, Ian McCulloch, Luca Tagliacozzo and Frank Verstraete. Support from the Australian Research Council, in the form of a Federation Fellowship (G. V.), is also acknowledged.

\appendix

\section{Algorithms for the evolutions in Fig.(\ref{TEBD1})}

In this appendix we explain in some detail how to implement the iTEBD algorithm for some particularly relevant choices of the gate $G$. In Sect. III we analyzed the case where the system is invariant under translations by $n=1$ lattice sites. Here we consider the case $n=2$ for the four different types of gate represented in Fig.(\ref{TEBD1}). Most of the information contained in the appendix can already be derived from the results of Sects. II and III, but we write it explicitly for each gate for the sake of clarity. 

\subsection{Matrix product operator, Fig.(\ref{TEBD1}.$i$)}
\begin{figure}
\includegraphics[width=0.5\textwidth]{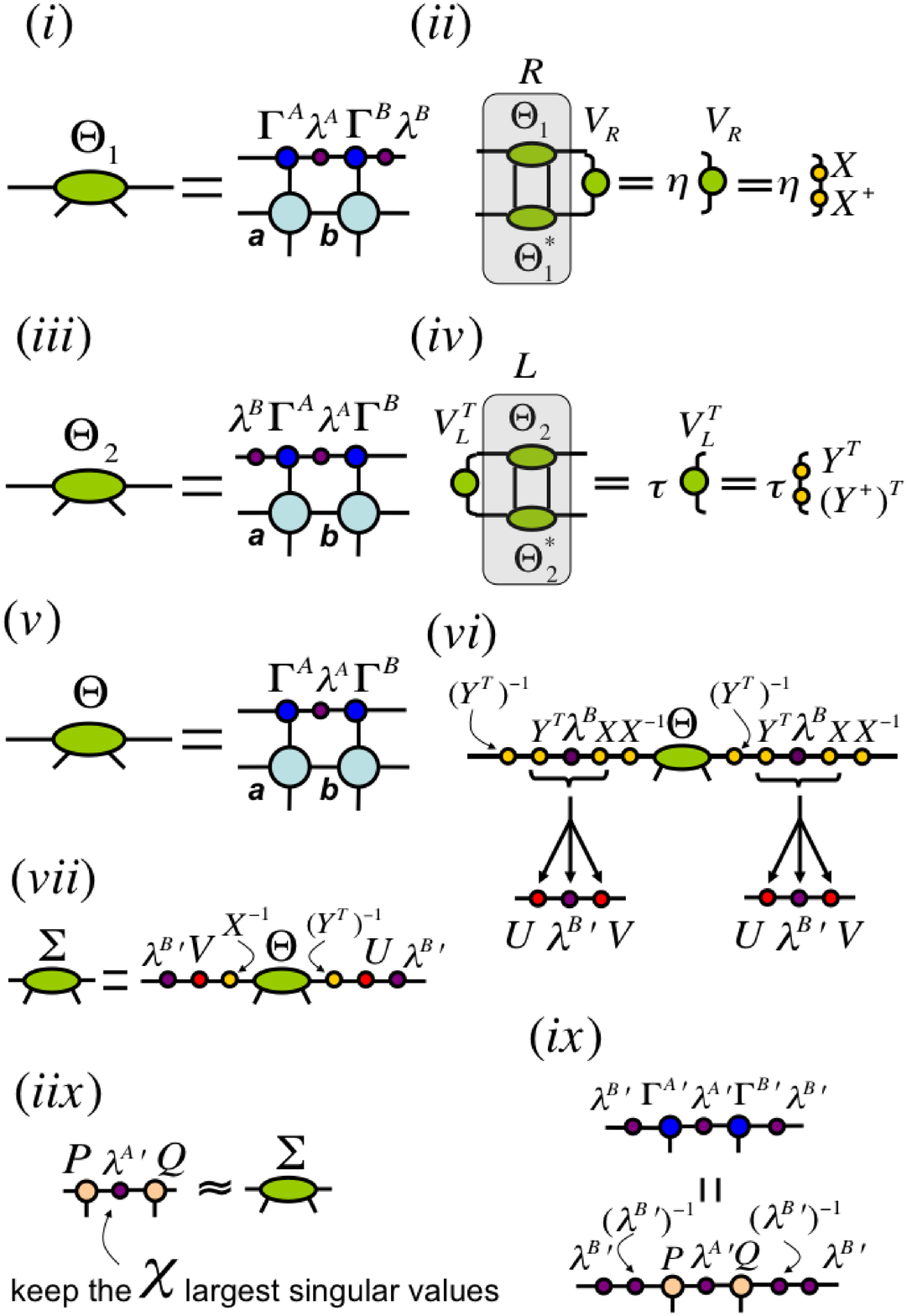}
\caption{(color online) Steps in the updating of the iMPS after the action of a MPO.}
\label{HVfig3pres}
\end{figure}

Let us assume that the iMPS for $\ket{\Psi}$ is given by some tensors $\Gamma^A,\lambda^A$ for odd sites and tensors $\Gamma^B,\lambda^B$ for even sites. We consider the evolution under a MPO invariant under translations of two chain sites,
\beq
G = \sum_{\alpha, \beta, \gamma, \delta, \rho, \ldots} \cdots a^{[i]}_{\alpha \beta} b^{[i+1]}_{\beta \gamma}a^{[i+2]}_{\gamma \delta} b^{[i+3]}_{\delta \rho} \cdots \ ,
\label{eqd}
\eeq
where for each value of $\alpha, \beta = 1,2, \ldots, \kappa$ the one-site operators $a^{[i]}_{\alpha \beta}:\mathcal{H}^{[i]} \rightarrow \mathcal{H}^{[i]}$ and $b^{[i+1]}_{\alpha \beta}:\mathcal{H}^{[i+1]} \rightarrow \mathcal{H}^{[i+1]}$ act on sites $i$ and $i+1$, see Fig.(\ref{TEBD1}.$i$). The updating procedure of the iMPS can be expressed as follows: 

$(i)$ Compute tensor $\Theta_1$ as indicated in Fig.(\ref{HVfig3pres}.$i$), with bond dimension $\kappa \chi$.

$(ii)$ Find the matrix $V_R$ that is the right dominant \cite{dominant} eigenvector of $R$ (in the sense of Fig.(\ref{HVfig3pres}.$ii$)) with dominant eigenvalue $\eta \in \mathbb{C}$ (assumed to be unique) \cite{eta}, where $R$ is obtained by contracting $\Theta_1$ with its complex conjugate $\Theta_1^*$ as shown in Fig.(\ref{HVfig3pres}.$ii$) [use large-scale, non-Hermitian eigenvalue solver, such as Arnoldi methods, and exploit the tensor network structure of $R$]. Then, decompose matrix $V_R$ (which is Hermitian and non-negative) as the square $V_R = X X^{\dagger}$. For instance, if $V_R = WDW^{\dagger}$ is the eigenvalue decomposition of $V_R$, then $X = W\sqrt{D}$.

$(iii)$ Compute tensor $\Theta_2$ as indicated in Fig.(\ref{HVfig3pres}.$iii$), with bond dimension $\kappa \chi$.

$(iv)$ Find the matrix $V_L$ that is the left dominant \cite{dominant} eigenvector of $L$ (in the sense of Fig.(\ref{HVfig3pres}.$iv$)) with dominant eigenvalue $\tau \in \mathbb{C}$ (assumed to be unique) \cite{eta}, where $L$ is obtained by contracting $\Theta_2$ with its complex conjugate $\Theta_2^*$ as shown in Fig.(\ref{HVfig3pres}.$iv$) [use large-scale, non-Hermitian eigenvalue solver, such as Arnoldi methods, and exploit the tensor network structure of $L$]. Then, decompose matrix $V_L$ (which is Hermitian and non-negative) as the square $V_L = Y^{\dagger} Y$, in the same way as was done for matrix $V_R$.

$(v)$ Compute tensor $\Theta$ as indicated in Fig.(\ref{HVfig3pres}.$v$), with bond dimension $\kappa \chi$.

$(vi)$ Introduce the two resolutions of the identity matrix $\mathbb{I} = (Y^T)^{-1} Y^T$ and $\mathbb{I} = XX^{-1}$ in the bond indices of tensor $\Theta$ as indicated in 
Fig.(\ref{HVfig3pres}.$vi$). Then, compute the singular value decomposition $Y^T X = U \lambda^{B \prime} V$, leading to new Schmidt coefficients $\lambda^{B \prime}$. Truncate these new Schmidt coefficients by keeping only the $\chi$ largest ones, and normalize them so that the sum of their squared values is 1.

$(vii)$ Compute tensor $\Sigma$ as indicated in Fig.(\ref{HVfig3pres}.$vii$).

$(iix)$ Group the indices of $\Sigma$ according to a single index for the left-hand side and a single index for the right-hand side, and compute the singular value decomposition as indicated in Fig.(\ref{HVfig3pres}.$iix$). This leads to two isometric tensors $P$ and $Q$, and new Schmidt coefficients $\lambda^{A \prime}$. Truncate these new Schmidt coefficients by keeping only the $\chi$ largest ones, and normalize them so that the sum of their squared values is 1.

$(ix)$ Obtain new matrices $\Gamma^{A \prime}$ and $\Gamma^{B \prime}$ as indicated in Fig.(\ref{HVfig3pres}.$ix$).

The above sequence of steps has a computational cost of $O(d^2 \kappa^3 \chi^3)$ in time. 

\subsection{Tensor product of two-site operators, Fig.(\ref{TEBD1}.$ii$)}

\begin{figure}
\includegraphics[width=0.5\textwidth]{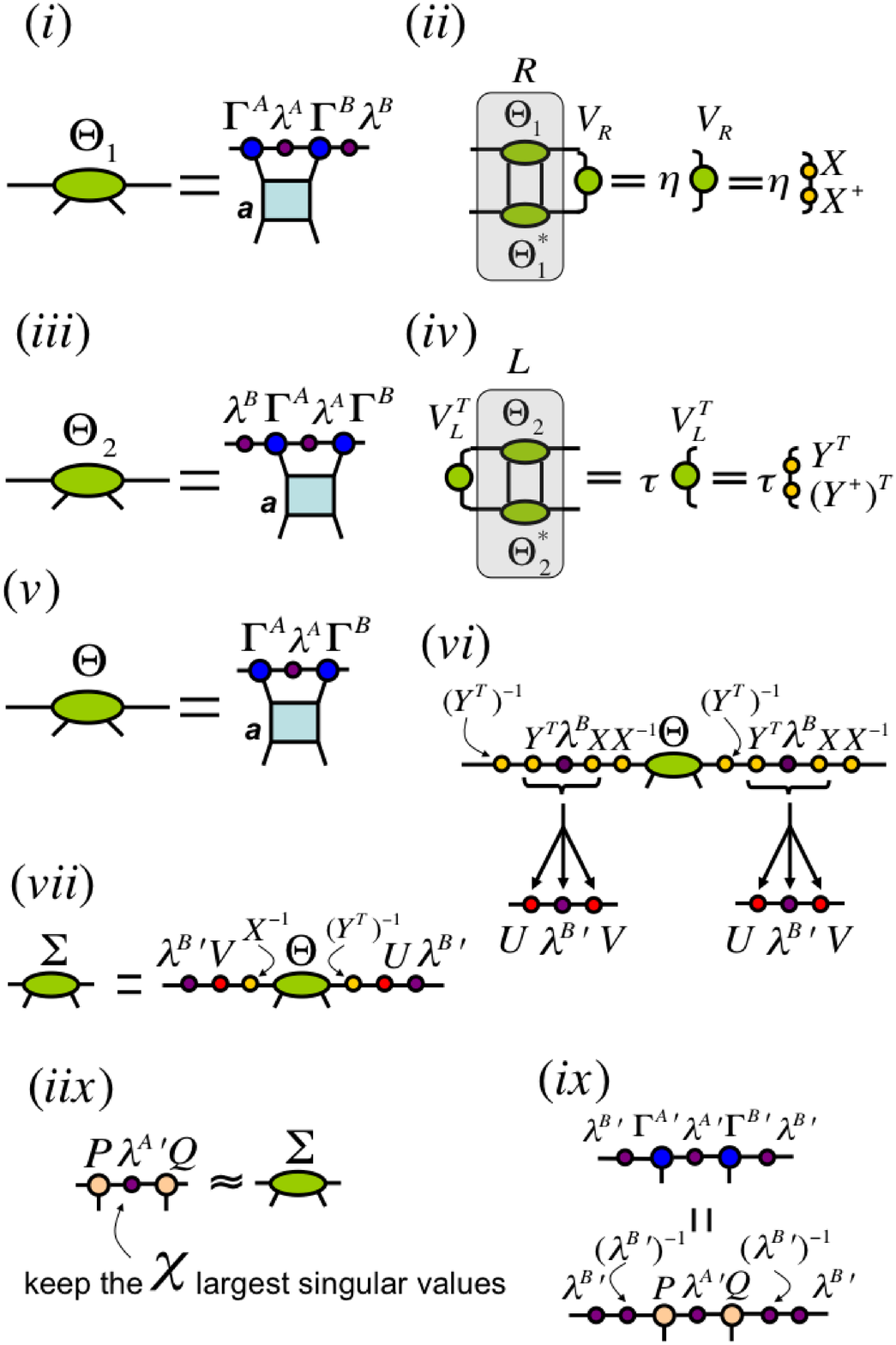}
\caption{(color online) Steps in the updating of the iMPS after the action of the tensor product of two-site operators.}
\label{fig3pres}
\end{figure}

Consider an iMPS for state $\ket{\Psi}$ with bond dimension $\chi$ that is invariant under shifts of two chain sites. This iMPS is then defined by tensors $\Gamma^A,\lambda^A$ for odd sites and tensors $\Gamma^B,\lambda^B$ for even sites. The evolution operator that we consider is given by
\beq
G = \bigotimes_{i \in \  {\rm odd}} a^{[i,i+1]} \ ,
\label{equ}
\eeq
where $a^{[i,i+1]}:\mathcal{H}^{[i]} \otimes \mathcal{H}^{[i+1]} \rightarrow \mathcal{H}^{[i]} \otimes \mathcal{H}^{[i+1]}$ is a two-body operator acting on the two contiguous sites $i$ and $i+1$ of the iMPS, see Fig.(\ref{TEBD1}.$ii$). The algorithm to update the iMPS is as follows:

$(i)$ Compute tensor $\Theta_1$ as indicated in Fig.(\ref{fig3pres}.$i$), with bond dimension $\chi$.

$(ii)$ Find the matrix $V_R$ that is the right dominant \cite{dominant} eigenvector of $R$ (in the sense of Fig.(\ref{fig3pres}.$ii$)) with dominant eigenvalue $\eta \in \mathbb{C}$ (assumed to be unique) \cite{eta}, where $R$ is obtained by contracting $\Theta_1$ with its complex conjugate $\Theta_1^*$ as shown in Fig.(\ref{fig3pres}.$ii$) [use large-scale, non-Hermitian eigenvalue solver, such as Arnoldi methods, and exploit the tensor network structure of $R$]. Then, decompose matrix $V_R$ (which is Hermitian and non-negative) as the square $V_R = X X^{\dagger}$. For instance, if $V_R = WDW^{\dagger}$ is the eigenvalue decomposition of $V_R$, then $X = W\sqrt{D}$.

$(iii)$ Compute tensor $\Theta_2$ as indicated in Fig.(\ref{fig3pres}.$iii$), with bond dimension $\chi$.

$(iv)$ Find the matrix $V_L$ that is the left dominant \cite{dominant} eigenvector of $L$ (in the sense of Fig.(\ref{fig3pres}.$iv$)) with dominant eigenvalue $\tau \in \mathbb{C}$ (assumed to be unique) \cite{eta}, where $L$ is obtained by contracting $\Theta_2$ with its complex conjugate $\Theta_2^*$ as shown in Fig.(\ref{fig3pres}.$iv$) [use large-scale, non-Hermitian eigenvalue solver, such as Arnoldi methods, and exploit the tensor network structure of $L$]. Then, decompose matrix $V_L$ (which is Hermitian and non-negative) as the square $V_L = Y^{\dagger} Y$, in the same way as was done for matrix $V_R$.

$(v)$ Compute tensor $\Theta$ as indicated in Fig.(\ref{fig3pres}.$v$), with bond dimension $\chi$.

$(vi)$ Introduce the two resolutions of the identity matrix $\mathbb{I} = (Y^T)^{-1} Y^T$ and $\mathbb{I} = XX^{-1}$ in the bond indices of tensor $\Theta$ as indicated in 
Fig.(\ref{fig3pres}.$vi$). Then, compute the singular value decomposition $Y^T X = U \lambda^{B \prime} V$, leading to new Schmidt coefficients $\lambda^{B \prime}$. Truncate these new Schmidt coefficients by keeping only the $\chi$ largest ones, and normalize them so that the sum of their squared values is 1.

$(vii)$ Compute tensor $\Sigma$ as indicated in Fig.(\ref{fig3pres}.$vii$).

$(iix)$ Group the indices of $\Sigma$ according to a single index for the left-hand side and a single index for the right-hand side, and compute the singular value decomposition as indicated in Fig.(\ref{fig3pres}.$iix$). This leads to two isometric tensors $P$ and $Q$, and new Schmidt coefficients $\lambda^{A \prime}$. Truncate these new Schmidt coefficients by keeping only the $\chi$ largest ones, and normalize them so that the sum of their squared values is 1.

$(ix)$ Obtain new matrices $\Gamma^{A \prime}$ and $\Gamma^{B \prime}$ as indicated in Fig.(\ref{fig3pres}.$ix$).

The computational cost of the above sequence of steps is $O(d^4 \chi^3)$. Also, and as expected, if $a^{[i,i+1]}$ is a unitary operator then this procedure corresponds exactly to the updating rules of the standard iTEBD algorithm.

\subsection{Tensor product of two-to-one-site operators, Fig.(\ref{TEBD1}.$iii$)}

\begin{figure}
\includegraphics[width=0.5\textwidth]{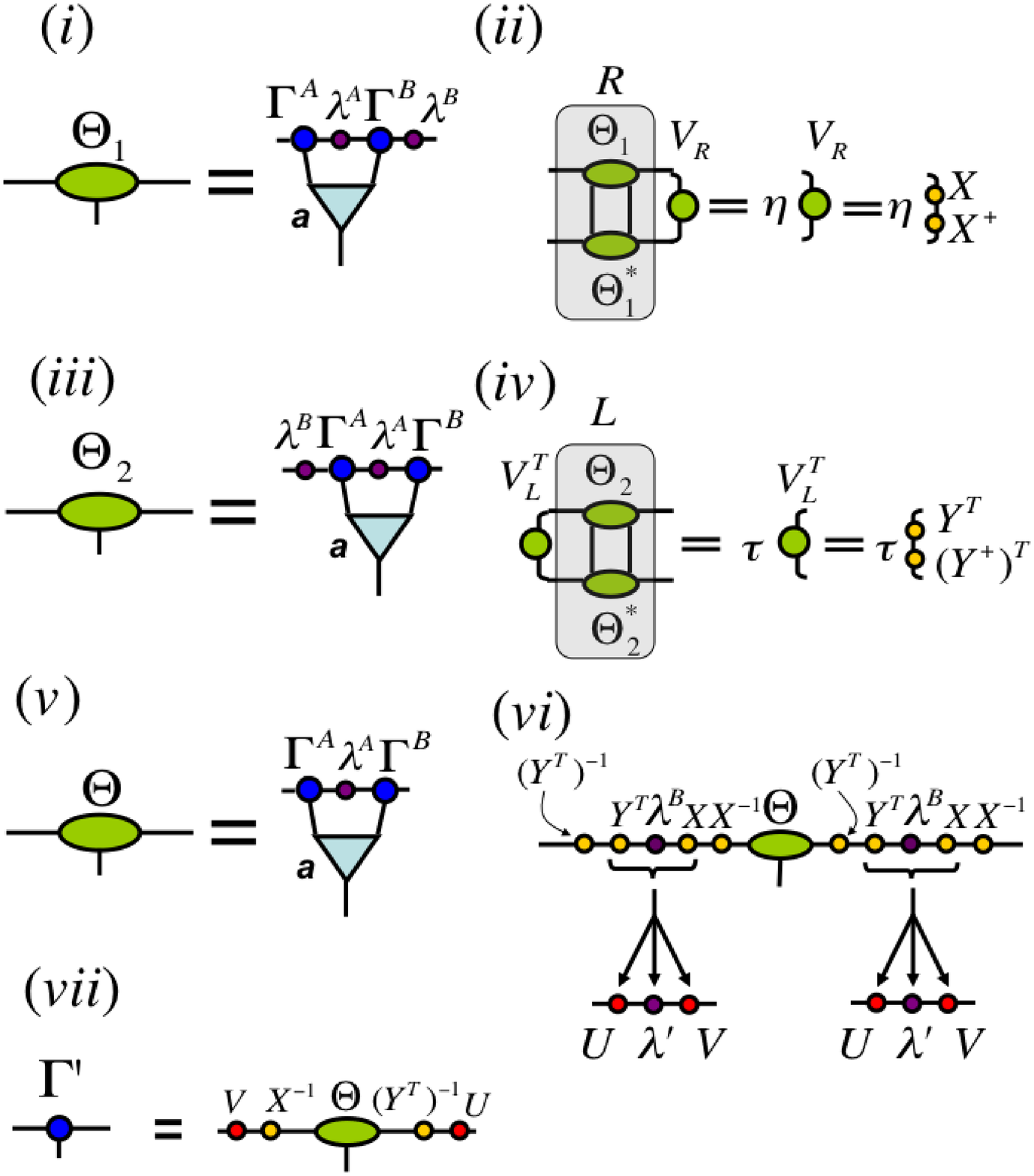}
\caption{(color online) Steps in the updating of the iMPS after the action of the tensor product of two-to-one-site gates.}
\label{twotoone}
\end{figure}

Our concern now is the evolution of an iMPS under an operator $G$ that is the tensor product of two-to-one-site operators
\beq
G = \bigotimes_{i \in \  {\rm odd}} a^{[i,i+1]} \ ,
\label{eqt}
\eeq
where $a^{[i,i+1]}:\mathcal{H}^{[i]} \otimes \mathcal{H}^{[i+1]} \rightarrow \mathcal{H}^{[(i+1)/2]}$ is a two-to-one-site operator acting on two contiguous sites $i$ and $i+1$ of the iMPS, and which maps the two sites to a new site $(i+1)/2$, see Fig.(\ref{TEBD1}.$iii$). Again we assume that the iMPS is defined by $\Gamma^A,\lambda^A$ for odd sites and $\Gamma^B,\lambda^B$ for even sites. The algorithm to update the iMPS is as follows:

$(i)$ Compute tensor $\Theta_1$ as indicated in Fig.(\ref{twotoone}.$i$), with bond dimension $\chi$.

$(ii)$ Find the matrix $V_R$ that is the right dominant \cite{dominant} eigenvector of $R$ (in the sense of Fig.(\ref{twotoone}.$ii$)) with dominant eigenvalue $\eta \in \mathbb{C}$ (assumed to be unique) \cite{eta}, where $R$ is obtained by contracting $\Theta_1$ with its complex conjugate $\Theta_1^*$ as shown in Fig.(\ref{twotoone}.$ii$) [use large-scale, non-Hermitian eigenvalue solver, such as Arnoldi methods, and exploit the tensor network structure of $R$]. Then, decompose matrix $V_R$ (which is Hermitian and non-negative) as the square $V_R = X X^{\dagger}$. For instance, if $V_R = WDW^{\dagger}$ is the eigenvalue decomposition of $V_R$, then $X = W\sqrt{D}$.

$(iii)$ Compute tensor $\Theta_2$ as indicated in Fig.(\ref{twotoone}.$iii$), with bond dimension $\chi$.

$(iv)$ Find the matrix $V_L$ that is the left dominant \cite{dominant} eigenvector of $L$ (in the sense of Fig.(\ref{twotoone}.$iv$)) with dominant eigenvalue $\tau \in \mathbb{C}$ (assumed to be unique) \cite{eta}, where $L$ is obtained by contracting $\Theta_2$ with its complex conjugate $\Theta_2^*$ as shown in Fig.(\ref{twotoone}.$iv$) [use large-scale, non-Hermitian eigenvalue solver, such as Arnoldi methods, and exploit the tensor network structure of $L$]. Then, decompose matrix $V_L$ (which is Hermitian and non-negative) as the square $V_L = Y^{\dagger} Y$, in the same way as was done for matrix $V_R$.

$(v)$ Compute tensor $\Theta$ as indicated in Fig.(\ref{twotoone}.$v$), with bond dimension $\chi$.

$(vi)$ Introduce the two resolutions of the identity matrix $\mathbb{I} = (Y^T)^{-1} Y^T$ and $\mathbb{I} = XX^{-1}$ in the bond indices of tensor $\Theta$ as indicated in 
Fig.(\ref{twotoone}.$vi$). Then, compute the singular value decomposition $Y^T X = U \lambda^{\prime} V$, leading to new Schmidt coefficients $\lambda^{\prime}$. Truncate these new Schmidt coefficients by keeping only the $\chi$ largest ones, and normalize them so that the sum of their squared values is 1.

$(ix)$ Obtain a new matrix $\Gamma^{\prime}$ as indicated in Fig.(\ref{twotoone}.$vii$).

The above procedure has a computational cost of $O(d^3 \chi^3)$. In the end, the action of the two-to-one-site gates $a^{[i]}$ on the iMPS can be computed exactly without further truncation of the bond indices, and is such that the obtained iMPS for the evolved state $\ket{\Psi'}$ is invariant under translations of one chain site instead of two.
\begin{figure}
\includegraphics[width=0.5\textwidth]{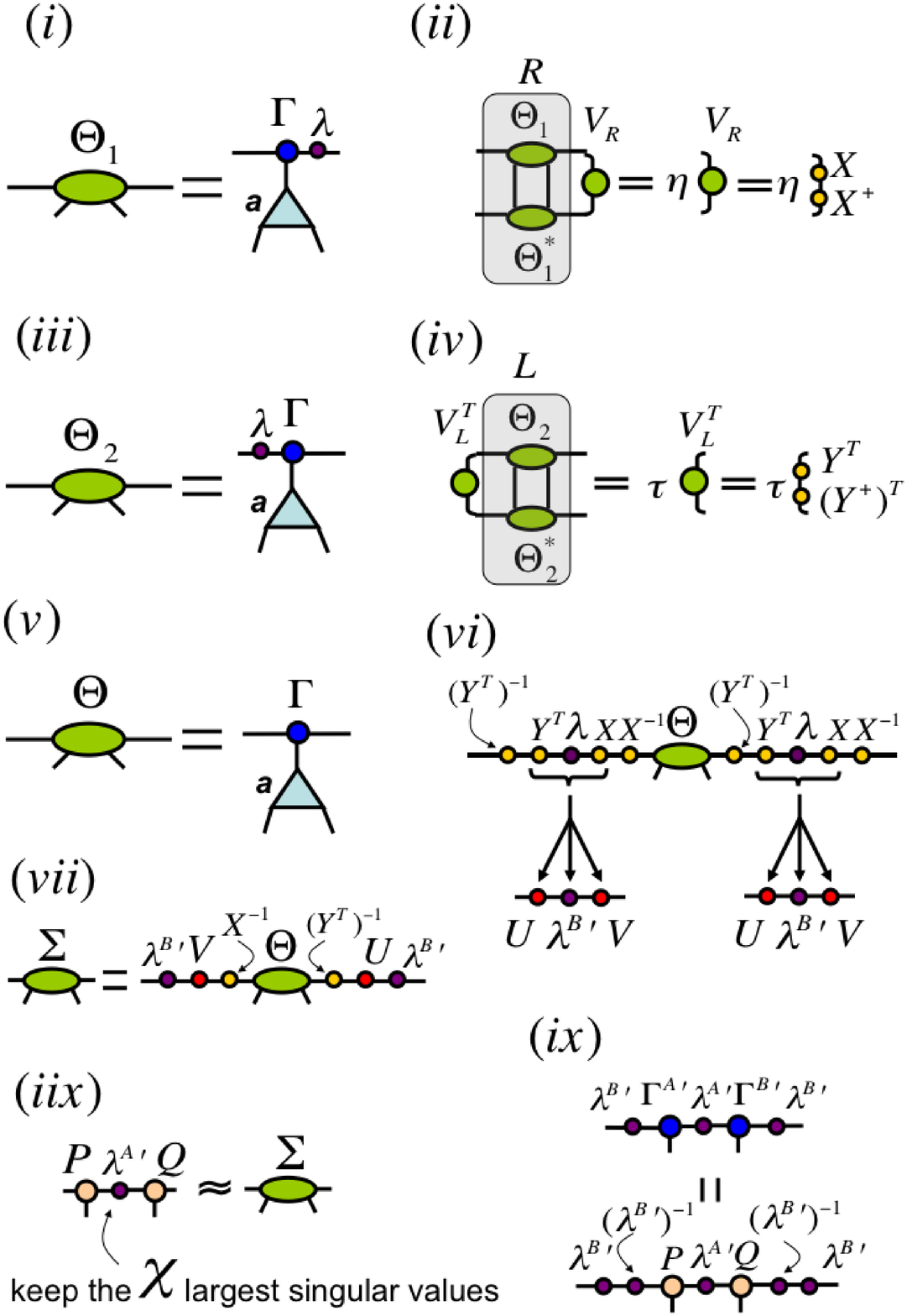}
\caption{(color online) Steps in the updating of the iMPS after the action of the tensor product of one-to-two-sites gates.}
\label{onetotwo}
\end{figure}

\subsection{Tensor product of one-to-two-sites operators, Fig.(\ref{TEBD1}.$iv$)}

Contrary to the previous cases, we consider now the situation in which the iMPS for state $\ket{\Psi}$ is defined by one tensor $\Gamma$ and one Schmidt vector $\lambda$, so that it is invariant under shifts of one chain site. At this point we wish to update the iMPS after the action of a tensor product of one-to-two-sites operators $G$ 
\beq
G = \bigotimes_{i} a^{[i]} \ ,
\label{eqq}
\eeq
where $a^{[i]}:\mathcal{H}^{[i]} \rightarrow \mathcal{H}^{[2i-1]} \otimes \mathcal{H}^{[2i]}$ is a one-to-two-sites operator acting on one site $i$ of the iMPS, and which maps the site to two new sites $2i-1$ and $2i$, see Fig.(\ref{TEBD1}.$iv$). The steps to follow to update the iMPS are: 

$(i)$ Compute tensor $\Theta_1$ as indicated in Fig.(\ref{onetotwo}.$i$), with bond dimension $\chi$.

$(ii)$ Find the matrix $V_R$ that is the right dominant \cite{dominant} eigenvector of $R$ (in the sense of Fig.(\ref{onetotwo}.$ii$)) with dominant eigenvalue $\eta \in \mathbb{C}$ (assumed to be unique) \cite{eta}, where $R$ is obtained by contracting $\Theta_1$ with its complex conjugate $\Theta_1^*$ as shown in Fig.(\ref{onetotwo}.$ii$) [use large-scale, non-Hermitian eigenvalue solver, such as Arnoldi methods, and exploit the tensor network structure of $R$]. Then, decompose matrix $V_R$ (which is Hermitian and non-negative) as the square $V_R = X X^{\dagger}$. For instance, if $V_R = WDW^{\dagger}$ is the eigenvalue decomposition of $V_R$, then $X = W\sqrt{D}$.

$(iii)$ Compute tensor $\Theta_2$ as indicated in Fig.(\ref{onetotwo}.$iii$), with bond dimension $\chi$.

$(iv)$ Find the matrix $V_L$ that is the left dominant \cite{dominant} eigenvector of $L$ (in the sense of Fig.(\ref{onetotwo}.$iv$)) with dominant eigenvalue $\tau \in \mathbb{C}$ (assumed to be unique) \cite{eta}, where $L$ is obtained by contracting $\Theta_2$ with its complex conjugate $\Theta_2^*$ as shown in Fig.(\ref{onetotwo}.$iv$) [use large-scale, non-Hermitian eigenvalue solver, such as Arnoldi methods, and exploit the tensor network structure of $L$]. Then, decompose matrix $V_L$ (which is Hermitian and non-negative) as the square $V_L = Y^{\dagger} Y$, in the same way as was done for matrix $V_R$.

$(v)$ Compute tensor $\Theta$ as indicated in Fig.(\ref{onetotwo}.$v$), with bond dimension $\chi$.

$(vi)$ Introduce the two resolutions of the identity matrix $\mathbb{I} = (Y^T)^{-1} Y^T$ and $\mathbb{I} = XX^{-1}$ in the bond indices of tensor $\Theta$ as indicated in 
Fig.(\ref{onetotwo}.$vi$). Then, compute the singular value decomposition $Y^T X = U \lambda^{B \prime} V$, leading to new Schmidt coefficients $\lambda^{B \prime}$. Truncate these new Schmidt coefficients by keeping only the $\chi$ largest ones, and normalize them so that the sum of their squared values is 1.

$(vii)$ Compute tensor $\Sigma$ as indicated in Fig.(\ref{onetotwo}.$vii$).

$(iix)$ Group the indices of $\Sigma$ according to a single index for the left-hand side and a single index for the right-hand side, and compute the singular value decomposition as indicated in Fig.(\ref{onetotwo}.$iix$). This leads to isometric two tensors $P$ and $Q$, and new Schmidt coefficients $\lambda^{A \prime}$. Truncate these new Schmidt coefficients by keeping only the $\chi$ largest ones, and normalize them so that the sum of their squared values is 1.

$(ix)$ Obtain new matrices $\Gamma^{A \prime}$ and $\Gamma^{B \prime}$ as indicated in Fig.(\ref{onetotwo}.$ix$). 

The computational cost of the above steps is $O(d^3 \chi^3)$. Similarly to the case of the previous section, the translational invariance of the original iMPS for $\ket{\Psi}$ has been modified, in a way that the obtained iMPS representation for the evolved state $\ket{\Psi'}$ has periodicity under shifts of two chain sites instead of one.


\begin{thebibliography}{100}

\bibitem{MPS}
S. Ostlund and S. Rommer, Phys. Rev. Lett. 75, 3537 (1995). M. Fannes, B. Nachtergaele, R. Werner, Commun. Math. Phys. 144, 443 (1992).  D. Perez-Garcia, F. Verstraete, M.M. Wolf, J.I. Cirac, Quantum Inf. Comput. 7, 401 (2007)

\bibitem{DMRG}
S. R. White, Phys. Rev. Lett. 69, 2863 (1992). S.R.White, Phys. Rev. B 48, 10345 (1992).

\bibitem{TEBD} 
G. Vidal, Phys. Rev. Lett. 91, 147902 (2003). G. Vidal, Phys. Rev. Lett. 93, 040502 (2004).  S. R. White, A. E. Feiguin, Phys. Rev. Lett. 93, 076401 (2004). A. J. Daley, C. Kollath, U. Schollwoeck, G. Vidal, J. Stat. Mech.: Theor. Exp. (2004) P04005. 

\bibitem{TPS}
T. Nishino, K. Okunishi, Y. Hieida, N. Maeshima, Y. Akutsu, Nucl. Phys. B 575 (2000) 504-512. T. Nishino, Y. Hieida, K. Okunishi, N. Maeshima, Y. Akutsu, A. Gendiar, Prog. Theor. Phys. 105 (2001) No.3, 409-417. A. Gendiar, N. Maeshima, T. Nishino, Prog. Theor. Phys. 110 (2003) No.4, 691-699. N. Maeshima, Y. Hieida, Y. Akutsu, T. Nishino, K. Okunishi, Phys. Rev. E64 (2001) 016705 [1-6]. Y. Nishio, N. Maeshima, A. Gendiar, T. Nishino, cond-mat/0401115. A. Gendiar, T. Nishino, R. Derian, Acta Phys. Slov. 55 (2005) 141. 

\bibitem{PEPS} 
F. Verstraete, J. I. Cirac, cond-mat/0407066. V. Murg, F. Verstraete, J. I. Cirac, Phys. Rev. A 75, 033605 (2007)

\bibitem{MERA} G. Vidal, Phys. Rev. Lett. 99, 220405 (2007). G. Vidal, arXiv:0707.1454 

\bibitem{TMRG}
T. Nishino, J. Phys. Soc. Jpn. 64 (1995) 3598-3601. T. Nishino, K. Okunishi, "Transfer-Matrix Approach to Classical Systems" Springer Lecture Note in Physics 528 ed. I. Peschel, X. Wang, K. Hallberg, Springer Berlin (1999) pp. 127-148. 

\bibitem{CTMRG}
T. Nishino, K. Okunishi, J. Phys. Soc. Jpn. 65, (1996) 891. T. Nishino, K. Okunishi, J. Phys. Soc. Jpn. 66, (1997) 3040. T. Nishino, K. Okunishi, M. Kikuchi, Physics Letters A 213, (1996) 69. 

\bibitem{TRG} M. Levin, C.P. Nave, Phys. Rev. Lett. 99, 120601 (2007). 

\bibitem{PWFRG}
T. Nishino, K. Okunishi, J. Phys. Soc. Jpn. 64 (1995) 4084-4087. K. Ueda, T. Nishino, K. Okunishi, Y. Hieida, R. Derian, A. Gendiar, J. Phys. Soc. Jpn. 75, 014003.1-014003.8 (2006). 

\bibitem{iDMRG}
I. P. McCulloch, arXiv:0804.2509.

\bibitem{iTEBD} G. Vidal, Phys. Rev. Lett. 98, 070201 (2007)

\bibitem{ZV} M. Zwolak and G. Vidal, Phys. Rev. Lett. 93, 207205 (2004).

\bibitem{VGC} F. Verstraete, J. J. Garcia-Ripoll, J. I. Cirac, Phys. Rev. Lett. 93, 207204 (2004).

\bibitem{dominant} Given a matrix $M$, we refer to its eigenvalue with largest absolute value as the \emph{dominant eigenvalue}. Similarly, we refer to the corresponding eigenvector as the \emph{dominant eigenvector} of $M$.

\bibitem{iPEPS}
J. Jordan, R. Or\'us, G. Vidal, F. Verstraete, J. I. Cirac, cond-mat/0703788; 

\bibitem{tTN} 
Y.-Y. Shi, L.-M. Duan and G. Vidal, Phys. Rev. A 74, 022320 (2006).

\bibitem{unique} The canonical form is unique up to a choice of phases $e^{i\phi_{\alpha}}$. Two canonical forms for $\ket{\Psi}$, $(\Gamma,\lambda)$ and $(\Gamma',\lambda')$, are related by  $(\Gamma')_{\alpha\beta}^i = e^{i\phi_{\alpha}} \Gamma_{\alpha\beta}^i e^{-i\phi_{\beta}}$ and $\lambda'=\lambda$.

\bibitem{eta} There are states of a chain, such as the cat state $\lim_{N\rightarrow \infty} c_0\ket{0}^{\otimes N}+c_1 \ket{1}^{\otimes N}$, $|c_0|^2 + |c_1|^2 =1$, for which the dominant eigenvalue is degenerate. The iMPS description needs to be supplemented with an extra tensor, sitting at infinite, that determines the boundary conditions (in this case the values of $c_0$ and $c_1$). Here we will not consider such cases.

\bibitem{Shi03} Non-Hermitian eigenvalue problems also occur in the context of transfer matrix DMRG. For instance, see N. Shibata, J. Phys. A: Math. and Gen. vol. 36 (2003) R381. 

\bibitem{Cho} A Cholesky decomposition can also be used to obtain two lower triangular matrices for $X$ and $Y$ (see e.g. {\it http://en.wikipedia.org/wiki/Cholesky$\_$decomposition}).

\bibitem{bound} This bound is optimal as long as $\chi > \kappa$. In the case $\kappa \ge \chi$ the efficiency can be improved by using alternative contractions.   

\bibitem{truncation} Truncating a bond index so as to retain the $\chi$ largest Schmidt coefficients $\lambda_{\alpha}$ is optimal in that it maximizes the overlap between the initial and truncated states. In the present case we use this recipe to truncate all bond indices of the iMPS at once. This is no longer expected to be optimal, but it is simple and seen to produce very satisfactory results.

\bibitem{compensation} Another way to turn an iMPS $\{\Gamma^A,\lambda^A,\Gamma^B,\lambda^B\}$ into the canonical form is by using the algorithm of Ref. \cite{iTEBD} to simulate a large sequence of trivial two-site gates (that is, gates that implement the identity operator) alternatively acting on even and odd bonds. It is seen that after each update the iMPS is closer to the canonical form. In practice, the orthonormalization strategy explained in this paper is more efficient and precise.

\bibitem{no_square_root} For a non-symmetric Hamiltonian $K_2$, $Q$ is decomposed into two different  matrices $Q=Q_1Q_2$ (e.g. through a singular value decomposition). If $K_2$ changes along different lattice directions (anisotropic model), then we will decompose two matrices $Q^{x}$ and $Q^{y}$. In both situations one can proceed in a similar way as in the symmetric, isotropic case.
 
\bibitem{PEPSclas} Our construction was inspired by a similar one in F. Verstraete, M. M. Wolf, D. Perez-Garcia, J. I. Cirac, Phys. Rev. Lett. 96, 220601 (2006), where finite systems were analyzed by mapping the partition function into a PEPS. Here we skip the map into PEPS and significantly reduce simulation costs by decreasing the bond dimension of the resulting 2D tensor network from $d^2$ to $d$.

\bibitem{converged} Another good reason to use the canonical form of an iMPS is that it simplifies the comparison between two states. As a criterion for convergence of the sequence $\ket{\Psi_p}$ in Eq. (\ref{eq:Psip}), we require that the Schmidt coefficients $\lambda$ of the iMPS have converged with respect to $p$ within some accuracy.  
 


%


\end{thebibliography}
\end{document}